\documentclass[epjCONF]{svjour}
\usepackage{graphicx}
\usepackage[varg]{txfonts} % Times fonts
\usepackage[latin1]{inputenc}
\session-title{FUSION11}
\begin{document}
\title{Beyond mean-field approach to heavy-ion reactions
around the Coulomb barrier}
\author{Kouhei Washiyama\inst{1}\fnmsep\thanks{\email{kouhei.washiyama@ulb.ac.be}} \and Denis Lacroix\inst{2} \and Sakir Ayik\inst{3} 
\and B\"ulent Yilmaz\inst{4}}
\institute{PNTPM, CP229, Universit\'e Libre de Bruxelles, 1050 Bruxelles, Belgium
\and GANIL, BP55027, 14076 Caen, France
\and Physics Department, Tennessee Technological University, Cookeville, 
Tennessee 38505, USA
\and Physics Department, Ankara University, Tandogan 06100, Ankara, Turkey}
\abstract{
Dissipation and fluctuations of one-body observables in heavy-ion reactions around
the Coulomb barrier are investigated with a microscopic
stochastic mean-field approach.
By projecting the stochastic mean-field dynamics on a suitable collective path,
transport coefficients associated with the relative distance between 
colliding nuclei and a fragment mass are extracted.
Although microscopic mean-field approach is know to underestimate
the variance of fragment mass distribution, the description of the variance 
is much improved by the stochastic mean-field method.
%The result is consistent with the macroscopic phenomenological 
%analysis of the experimental data.
While fluctuations are consistent with the empirical (semiclassical) analysis
of the experimental data, concerning mean values of macroscopic variables
the semiclassical description breaks down below the Coulomb barrier.
} %end of abstract
\maketitle
\section{Introduction}
\label{intro}
The interplay between nuclear structure and dynamical effects is crucial to 
properly describing heavy-ion fusion reactions at energies close to 
the Coulomb barrier.
%Coupled-channels models have been widely used to describe
%the entrance channel of fusion reactions. 
The mean-field theory based on the Skyrme energy density functional 
provides a rather unique tool for describing nuclear structure and 
nuclear reactions over the whole nuclear chart in a unified framework.
For nuclear dynamical calculations, time-dependent Hartree-Fock (TDHF) model
has been developed~\cite{bonche76,koonin80,negele82,Sim08}.
This model automatically includes important dynamical effects 
such as vibration of nuclei, neck formation, and nucleon transfer 
during reactions. Most recent TDHF simulations are able to include all terms of 
the Skyrme energy density functional, for example the spin-orbit terms, 
%which has been already included in static calculations 
and to break spacial symmetries~\cite{kim97,simenel01,nakatsukasa05,maruhn05,umar06b,iwata10}.
To perform full three-dimensional calculations and to use effective forces 
consistent with static calculations is crucial to accounting for 
the richness of nuclear shapes in dynamical evolution.

In the mean-field theory, short range two-body correlations are neglected and 
nucleons move in the self-consistent potential produced by all other nucleons. 
This is a good approximation at low energies since Pauli blocking is 
effective for scattering into unoccupied states. Consequently, 
collective energy is converted into intrinsic degrees of freedom 
via interaction of nucleons with the self-consistent mean field, 
so-called one-body dissipation. One-body dissipation mechanism plays 
a dominant role in low energy nuclear dynamics.
One important limitation of the mean-field theory is related with dynamical 
fluctuations of collective variables. In the mean-field description, 
while single-particle motion is treated in a quantal framework, 
collective motion is treated almost in a classical approximation. 
Therefore, TDHF provides a good description for average evolution; 
however it severely underestimates fluctuations of collective variables~\cite{koonin77,davies78,dasso79}.

After the first application of the TDHF theory, 
much effort has been devoted to overcoming this difficulty and 
to developing transport theories that are able to describe not only mean values 
but also fluctuations (for a review, see Refs~\cite{Abe96,Lac04}). 
Among them, the variational principle by Balian and V\'en\'eroni
appears as one of the most promising methods~\cite{Bal84,Mar85}. 
However, even nowadays it remains difficult to apply~\cite{broomfield08,simenel11}. 
%More than 30 years after the first application of the TDHF theory, 
The absence of a practical solution to include fluctuations beyond mean field 
in a fully microscopic framework strongly restricts applications of 
mean-field-based theories to low energy nuclear reactions.

A stochastic mean-field (SMF) approach has been proposed for describing
fluctuation dynamics~\cite{ayik08}. For small amplitude fluctuations, 
this model gives a result for dispersion of a one-body observable 
that is identical to the result obtained through a variational 
approach~\cite{Bal84}. 
%It is also shown that,
%when the SMF evolution is projected on a collective variable, it gives rise to 
%a generalized Langevin equation, which incorporates one-body dissipation and
%one-body fluctuation mechanisms in accordance with quantal
%dissipation-fluctuation relation. These illustrations give a strong
%support that the SMF approach provides a consistent microscopic description
%for the dynamics of density fluctuations in low energy nuclear reactions. 
 
In this contribution, 
first, we discuss the property of nucleus-nucleus potential
and one-body dissipation deduced from TDHF calculations.
They are not constrained by adiabatic or diabatic approximation.
Therefore, this method should provide an accurate description of 
nucleus-nucleus potential and one-body dissipation at energies
around the Coulomb barrier~\cite{washiyama08,washiyama09a}.
Then, we discuss how to overcome the failure of
description of fluctuations of one-body observables in TDHF and 
employ the SMF approach~\cite{ayik08,ayik09}.
%This approach is a stochastic extention of the TDHF model
%for low energy nuclear dynamics so as to include initial fluctuations 
%in collective space.
%The initial fluctuations are simulated by a suitable ensemble
%of initial single-particle density matrices.
We project the SMF evolution on a collective path to obtain the expression of
diffusion coefficients.
%
%By including initial fluctuations in collective space, 
We show that the description of the variance of fragment mass distributions
in transfer reactions by the stochastic mean-field approach is much improved 
compared to that of the TDHF model~\cite{washiyama09b}.

\section{Nucleus-nucleus potential and one-body dissipation from mean-field dynamics}
\label{sec:potential}

Nucleus-nucleus potential and one-body dissipation
are extracted as follows~\cite{washiyama08,washiyama09a}:
(i)~We solve the TDHF equation for head-on collisions 
to obtain the time evolution of the total density of colliding nuclei.
(ii)~Defining a separation plane between two nuclei, 
we compute at each time the relative distance $R$,
associated momentum $P$, and reduced mass $\mu$ of colliding nuclei.
(iii)~We assume that mean-field evolution obeys a classical equation of motion 
with a friction term:
\begin{eqnarray}
\frac{dR}{dt}=\frac{P}{\mu },~~~~
\frac{dP}{dt}=-\frac{dV}{dR}-\gamma (R)\dot{R},
\label{eq:newtonequation}
\end{eqnarray}
where $V(R)$ and $\gamma (R)$ denote the nucleus-nucleus potential 
and friction coefficient, respectively. The friction coefficient $\gamma (R)$ 
describes the effect of one-body energy dissipation from 
the macroscopic collective degrees of freedom to the microscopic ones.
For the TDHF calculations presented in this contribution, 
the three-dimensional TDHF code developed by P.~Bonche 
and coworkers with the SLy4d Skyrme effective force~\cite{kim97} is used.
The mesh sizes in space and in time are 0.8~fm and 0.45~fm/$c$, respectively.
For more details, see Refs.~\cite{washiyama08,washiyama09a}.

\begin{figure}[bthp]
\begin{center}\leavevmode
\resizebox{0.95\columnwidth}{!}{%
  \includegraphics{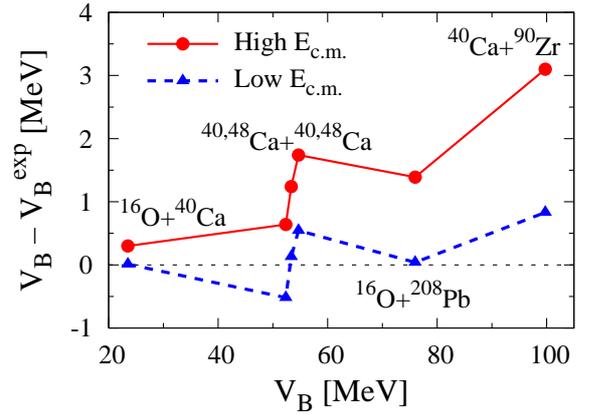} }
\caption{
Barrier height $V_B$ deduced from TDHF minus experimental barrier 
height $V_B^{\rm exp}$ as a function of extracted barrier height 
for the reactions indicated in the figure. 
The values $V_B$ are deduced from high energy (solid line) 
and from low energy ($E_{\rm c.m.}\sim V_B$) (dashed line) TDHF trajectories, respectively.
}
\label{fig:systematics}
\end{center}
\end{figure}

%Dynamical effect on potentials deduced from TDHF trajectories at 
%center-of-mass energies close to the Coulomb barrier is seen 
%in all reactions considered here. 
Figure~\ref{fig:systematics} shows the difference between the barrier height 
deduced from TDHF evolution ($V_B$) and the experimental barrier height 
($V_B^{\rm exp}$)~\cite{newton04} as a function of $V_B$. 
The solid line corresponds to the barrier height extracted using 
high-energy TDHF trajectories ($E_{\rm c.m.}\gg V_B$),
whereas the dashed line is the result for low-energy TDHF trajectories 
($E_{\rm c.m.}\sim V_B$).
The former identifies with the barrier height of 
the frozen density approximation~\cite{denisov02}. 
Dynamical reduction of the barrier height from high-energy TDHF 
to low-energy TDHF is clearly seen for all reactions. 
Moreover, because of this reduction, 
the value of the barrier height at low energy approaches the experimental data. 
This underlines the importance of dynamical effects close to the Coulomb barrier
and shows the precision of our method.

\begin{figure}[tbhp]
\begin{center}\leavevmode
\resizebox{0.95\columnwidth}{!}{%
  \includegraphics{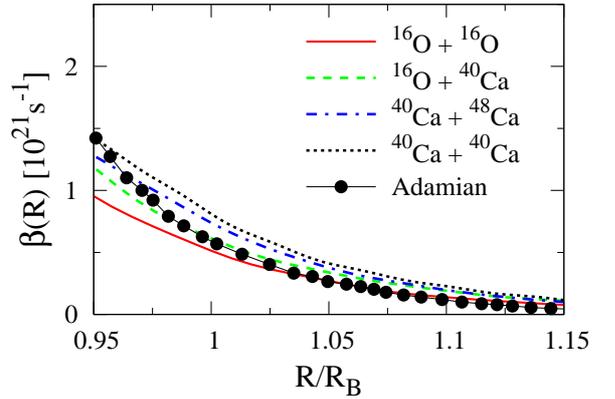} }
\caption{
Extracted reduced friction $\beta(R)\equiv\gamma(R)/\mu $ as a function of $R$ 
scaled by the Coulomb barrier radius $R_B$ for different systems. 
A microscopic friction coefficient 
by Adamian {\it et al.}~\cite{adamian97} is shown by the solid circles for comparison.
}
\label{fig:friction}
\end{center}
\end{figure}

This method also provides information on one-body dissipation
through the friction coefficient $\gamma$. 
In Fig.~\ref{fig:friction}, we present reduced friction coefficients 
$\beta (R)\equiv\gamma (R)/\mu $ as a function of $R$ 
scaled by the Coulomb barrier radius $R_B$ for different systems.
This clearly shows that the order of magnitude of 
$\beta(R)$ and the radial dependence are almost independent 
on the size and asymmetry of the system.
We compare these friction coefficients with that of a microscopic model 
based on small amplitude response by Adamian {\it et al.}~\cite{adamian97} 
by the solid circles. The radial dependence and the magnitude of 
their friction coefficient are very similar to those extracted from TDHF evolution.

\section{Fluctuations of one-body observables from stochastic mean-field dynamics}

In the previous section, we show that mean-field dynamics gives a
good description for the nucleus-nucleus potential and one-body dissipation.
However, it is well known that TDHF can not reproduce fluctuations
of one-body observables, for example the variance of fragment
mass distributions in deep inelastic collisions~\cite{dasso79},
%It has been recognized for a long time that TDHF
%calculations severely underestimate the experimental data of 
%the dispersion of mass distributions, 
although TDHF calculations well reproduce the mean value of fragment mass.

In order to overcome this difficulty, 
we have recently proposed a stochastic mean-field (SMF) approach, 
which is a stochastic extension of the mean-field model for 
low energy nuclear dynamics so as to include zero-point fluctuations of 
the initial state~\cite{ayik08,ayik09}. The initial density fluctuations 
are simulated by representing the initial state in terms of 
a suitable ensemble of initial single-particle density matrices,
which is similar to the idea in Refs.~\cite{esbensen78,dasso92}.
%In fact, this idea can be regarded as the beginning of 
%constructing time-dependent version of configuration mixing calculations.
%
In this manner, the description with a single Slater determinant is replaced by
a superposition of several Slater determinants. A member of the ensemble
of density matrices, indicated by event label $\lambda$, can be expressed as
\begin{eqnarray}
\label{eq:density}
\rho^\lambda({\bf r},{{\bf r}}^{\prime},t) = \sum\limits_{ij \sigma\tau}
\Phi_{i \sigma\tau}^\ast({\bf r},t;\lambda )\rho_{ij}^\lambda(\sigma\tau)
\Phi_{j \sigma\tau}({\bf r}^{\prime},t;\lambda ),
\end{eqnarray}
where the sums $i$ and $j$ run over a complete set of single-particle 
wave functions $\Phi_{i \sigma\tau}({\bf r},t;\lambda )$
with spin-isospin quantum numbers $\sigma$, $\tau$.
According to the description of the SMF approach~\cite{ayik08},
matrix elements $\rho_{ij}^\lambda(\sigma\tau)$
are assumed to be time-independent random Gaussian numbers with a mean value
$\overline{\rho_{ij}^\lambda(\sigma\tau)}=\delta_{ij}n_i^{\sigma\tau}$
and with a variance
\begin{eqnarray}
&&\overline{\delta\rho_{ij}^\lambda(\sigma\tau)
\delta\rho_{j'i'}^\lambda( {\sigma}'{\tau}')}\nonumber \\
&=&\frac{1}{2}\delta_{j{j}'}\delta_{i{i}'}\delta_{\tau {\tau}'}\delta_{\sigma {\sigma}'}
\left[n_i^{\sigma\tau}(1 - n_j^{\sigma\tau}) + n_j^{\sigma\tau}(1 - n_i^{\sigma\tau})\right],
\label{variance}
\end{eqnarray}
where $\overline{X}$ denotes the ensemble average of $X$.
Here, $n_i^{\sigma\tau}$ denotes the average single-particle occupation factor. At zero temperature
occupation factors are $0$ and $1$, and at finite temperature they are determined 
by the Fermi-Dirac distribution.
The great advantage of the SMF approach is that each Slater determinant
evolves independently from each other following the time evolution
of its single-particle wave functions in its self-consistent mean-field Hamiltonian, 
denoted by $h(\rho^\lambda)$, according to
\begin{eqnarray}
\label{eq:spwf}
i\hbar \frac{\partial }{\partial t}\Phi_{i \sigma\tau} ({\bf r},t;\lambda )
= h(\rho^\lambda )\Phi_{i \sigma\tau}({\bf r},t;\lambda ).
\end{eqnarray}

In the following applications, we focus on the head-on %$^{40}$Ca+$^{40}$Ca 
collision along the $x$ axis around the Coulomb barrier energy.

\subsection{Fusion reactions}

First, we apply the SMF approach to fusion reactions~\cite{ayik09}. 
%To discuss the fluctuation of collective variables,
We project the SMF evolution on a one-dimensional macroscopic Langevin equation,
which is similar to Eq.~(\ref{eq:newtonequation}) except
an additional Gaussian random force $\xi_P^\lambda(t)$: 
\begin{eqnarray}
\label{eq:langevin} \frac{d}{dt}P^{\lambda} = - \frac{d}{dR^{\lambda}}U(R^{\lambda} ) 
- \gamma (R^{\lambda} )\dot {R}^{\lambda} + \xi_P^{\lambda} (t),
\end{eqnarray}
Ignoring non-Markovian effects, the random force $\xi_P^\lambda(t)$ with zero mean value 
reduces to white noise specified by the following correlation function:
\begin{eqnarray}
\label{eq:correlation}
\overline{\xi _P^\lambda (t)\xi _P^\lambda({t}')} = 2\delta(t-{t}')D_{PP}(R).
\end{eqnarray}
Here $D_{PP}(R)$ denotes the momentum diffusion coefficient.
Denoting $x_0$ as the position of the separation plane 
between the two nuclei ($x_0=0$ in this case),
we have the following semiclassical expression for the nucleon 
diffusion coefficient according to Ref.~\cite{ayik09}:
\begin{figure}[tbph]
\begin{center}\leavevmode
\resizebox{0.95\columnwidth}{!}{%
  \includegraphics{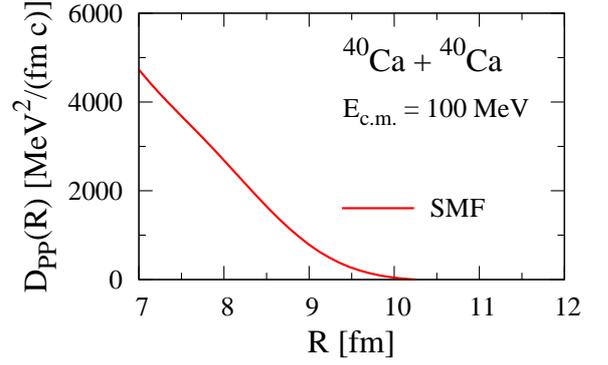} }
\caption{
Diffusion coefficient as a function of the relative distance
for the head-on $^{40}$Ca+$^{40}$Ca collision at $E_{\rm c.m.}=100$~MeV.
}
\label{fig:Dpp}
\end{center}
\end{figure}
\begin{eqnarray}
D_{PP}(t) &=& \int\frac{dp_x }{2\pi \hbar}
\frac{\vert p_x \vert}{m}\frac{p_x^2}{2} \nonumber \\
&&{}\times\sum_{\sigma\tau} \bigl\{f_P^{\sigma\tau}(x_0,p_x ,t)
\left[1 - f_T^{\sigma\tau}(x_0,p_x,t)\right/\Omega(x_0,t)] \nonumber \\
&&{}+ f_T^{\sigma\tau}(x_0,p_x,t)\left[1 - f_P^{\sigma\tau}(x_0,p_x,t)\right]/\Omega(x_0,t)\bigr\}.
\label{eq:DPP}\end{eqnarray}
Here $m$ is the nucleon mass,  
$\Omega(x_0,t)$ is the phase-space volume over the window 
given in Ref.~\cite{ayik09}, and
\begin{eqnarray}
 &&f_{P/T}^{\sigma\tau}(x_0,p_x ,t) \nonumber \\
&=& \iint dydz
\int ds_x\exp\left( - \frac{i}{\hbar}p_x s_x\right) \nonumber \\
&&{}\times  \sum_{i\in P/T}
\Phi_{i \sigma\tau}^\ast \left(x + \frac{s_x}{2},y,z,t\right)
n_{i}^{\sigma\tau} \Phi_{i \sigma\tau}\left(x - \frac{s_x}{2},y,z,t\right)
\nonumber \\
\label{eq:wignerPT}
\end{eqnarray}
is the average value of the reduced Wigner distribution
associated with single-particle wave functions originating from
the projectile/target. 
We note that the expression of the diffusion coefficient~(\ref{eq:DPP})
has the same form as that given by the phenomenological nucleon exchange model 
in Ref. \cite{feldmeier87}. We also note that diffusion coefficients 
can be evaluated in terms of the average TDHF evolution 
through the Wigner transformation~(\ref{eq:wignerPT}). 

As an example, the momentum diffusion coefficient for the head-on 
$^{40}$Ca+$^{40}$Ca collision at $E_{\rm c.m.}=100$~MeV is shown in 
Fig.~\ref{fig:Dpp}.

\subsection{Variance of fragment mass distribution in transfer reactions}

We show another application of SMF to the variance of fragment 
mass distribution in transfer reactions 
to improve the mean-field description~\cite{washiyama09b}.
To do so, we investigate head-on transfer reactions at energies just below 
the Coulomb barrier, where nucleon exchange will occur during reaction,
and estimate the variance of fragment mass distribution.

By projecting the SMF evolution on the collective space for the fragment
mass number, time evolution of the mass number of the projectile-like 
fragment $A_P^\lambda $ for an event $\lambda$ 
is described by a Langevin equation~\cite{Randrup2},
\begin{eqnarray}
\label{eq:langevin-mass}
\frac{d}{dt}A_P^\lambda = v(A_P^\lambda, t) + \xi_A^\lambda(t),
\end{eqnarray}
where $v(A_P^\lambda ,t)$ denotes the drift term for nucleon transfer.
The Gaussian white noise random force $\xi_A^\lambda (t)$ is determined 
with zero mean value and a correlation function,
\begin{equation}
\overline{\xi_A^\lambda (t)\xi_A^\lambda ({t}')}= 2\delta(t -{t}')D_{AA},
\end{equation}
where $D_{AA}$ is the diffusion coefficient associated with nucleon exchange.
The variance $\sigma_{AA}^{2}$ of fragment mass distribution is determined 
by small fluctuations of the mass number $\delta A_{P}^{\lambda}$ through 
$\sigma_{AA}^{2}(t)=\overline{\delta A_{P}^{\lambda}\delta A_{P}^{\lambda}}$. 
The expression of the diffusion coefficient $D_{AA}$ is similar to 
the momentum diffusion coefficient $D_{PP}$ except that 
$p_x^2$ is removed from the integral in  Eq.~(\ref{eq:DPP}).
Again, we note that the nucleon diffusion coefficient can also be evaluated 
from the average TDHF evolution.

According to the Langevin equation, neglecting
contributions from the drift term, the variance is related to the diffusion
coefficient according to \cite{Randrup2,Randrup82}
\begin{eqnarray}
\sigma^2_{AA}(t) \simeq 2 \int_0^t D_{AA}(s)ds.
\label{eq:sigma}
\end{eqnarray}
In the phenomenological nucleon exchange model, 
the relation $ \sigma_{AA}^{2}(t)=N_{\rm exc}(t)$ was obtained,
where $N_{\rm exc}(t)$ denotes the accumulated total number of exchanged nucleons until time $t$,
and was extensively used to analyze the experimental data of mass variance~\cite{Fre84,Ada94}.
In the following, to check whether the SMF approach satisfies the above relation,
we estimate the both quantities by the SMF approach.

\begin{figure}[bthp]
\begin{center}\leavevmode
\resizebox{0.95\columnwidth}{!}{%
  \includegraphics{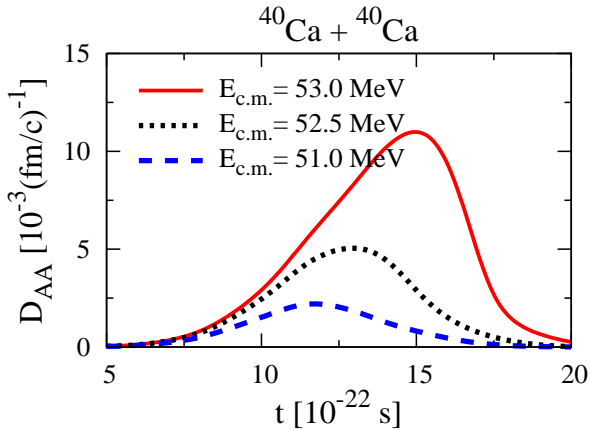} }
\caption{Time evolution of diffusion coefficient calculated 
in the SMF approach for the head-on
$^{40}{\rm Ca}+{}^{40}{\rm Ca}$ reaction at different center-of-mass energies
below the Coulomb barrier.
}
\label{fig:DAAtime}
\end{center}
\end{figure}

We carry out calculations for the head-on $^{40}$Ca+$^{40}$Ca reaction
at energies just below the Coulomb barrier.
Figure~\ref{fig:DAAtime} illustrates the dependence of diffusion coefficients 
at different center-of-mass energies. 
The Coulomb barrier energy of this system is 53.4 MeV.
\begin{figure}[bthp]
\begin{center}\leavevmode
\resizebox{0.95\columnwidth}{!}{%
  \includegraphics{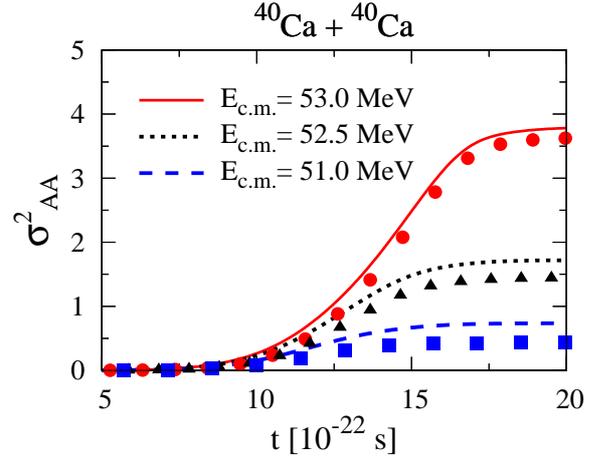} }
\caption{
Time evolution of $\sigma_{AA}^2$ obtained from the SMF approach for the 
$^{40}$Ca+$^{40}$Ca reaction at the same energies as those in Fig.~\ref{fig:DAAtime}.
Number of exchanged nucleon is superimposed by the solid circles, 
solid triangles, and solid squares from high to low energies, respectively.
}
\label{fig:SIGMAtime}
\end{center}
\end{figure}
The magnitude of diffusion coefficient essentially depends on the size of 
the window, the larger the window the larger the rate of change of 
nucleon exchange~\cite{Randrup82}. 
At a given center-of-mass energy, the diffusion coefficient becomes maximum 
at the turning point where the size of the window is the largest.  
Also, because of increasing overlap of the projectile and the target, 
the magnitude of the diffusion coefficient increases with energy.

In Fig.~\ref{fig:SIGMAtime},
the variance of the fragment mass distributions deduced from the SMF approach 
at the same energies as those in Fig.~\ref{fig:DAAtime} are shown by lines.
The corresponding evolution of the number of exchanged nucleons is 
superimposed by the solid circles, solid triangles, and solid squares 
from high to low energies, respectively.
The mass variance estimated from the SMF approach is consistent 
with this relation. We also estimate the variance of fragment mass 
distribution using the standard TDHF approach. 
%The asymptotic values of $\sigma_{AA}^2$ for the $^{40}$Ca+$^{40}$Ca reaction 
%are 0.004, 0.008, and 0.008 from low to high energies, 
%while the number of exchanged nucleons are 0.432, 1.441, and 3.634.
%
In Table~1, the asymptotic values of the variances obtained from SMF and 
from TDHF are compared with those of the number of exchanged nucleons.
The TDHF results are much smaller than the number of exchanged nucleons and
are also much smaller than the results obtained from the SMF approach.
\begin{table}[bthp]
\caption{Asymptotic values of the variances obtained from 
TDHF ($\sigma^2_{\rm TDHF}$) and SMF ($\sigma_{AA}^2$) for the 
$^{40}$Ca+$^{40}$Ca reaction.
Asymptotic values of the number of exchanged nucleons
are also given in the last column.}
\label{tab:1}       % Give a unique label
% For LaTeX tables use
\begin{center}
\begin{tabular}{llll}
\hline\noalign{\smallskip}
$E_{\rm c.m.}$ & $\sigma^2_{\rm TDHF}$ & $\sigma_{AA}^2$  &$N_{\rm exc}$  \\
\noalign{\smallskip}\hline\noalign{\smallskip}
51.0 & 0.004 & 0.730 & 0.432 \\
52.5 & 0.004 & 1.718 & 1.441 \\
53.0 & 0.008 & 3.790 & 3.634 \\
\noalign{\smallskip}\hline
\end{tabular}\end{center}
\end{table}
The failure of the TDHF theory on the description of the variance of 
the fragment mass distribution has been recognized for a long time 
as a major limitation of the mean-field theory~\cite{koonin77,davies78,dasso79}.
It appears that 
the SMF approach cures this shortcoming of the mean-field theory.
As seen from Fig.~\ref{fig:SIGMAtime}, 
not only the asymptotic value of $\sigma^2_{AA}$ but also the entire
time evolution is very close to the evolution of $N_{\rm exc}(t)$.

%Before closing this section, we would like to show 
%in Fig.~\ref{fig:SIGMAtimeCaZr} a preliminary result
%of the time evolution of the variance $\sigma_{AA}^2$
%for the $^{40}$Ca+$^{90}$Zr reaction
%by extending the SMF approach to the asymmetric system
%including the motion of the neck position~\cite{bulent}.
%Again, good agreement between the variance and the number of exchanged nucleon
%is obtained, which is further strong support of the SMF approach.

%\begin{figure}[bthp]
%  \includegraphics[width=0.95\columnwidth,clip]{sigmaCaZr.eps} 
%\caption{Time evolution of $\sigma_{AA}^2$ by the solid line 
%obtained from the SMF approach is compared with the number of exchanged 
%nucleon by the solid circles for the $^{40}$Ca+$^{90}$Zr reaction.}
%\label{fig:SIGMAtimeCaZr}
%\end{figure}

\section{Nucleon drift in asymmetric system}

%We would like to add a few comments on SMF applications to asymmetric systems,
The work presented in the previous section has been extended to 
asymmetric systems. Then, in addition to the diffusion coefficient, 
the drift term in Eq.~(\ref{eq:langevin-mass}) connected to 
the average nucleon transfer is different from zero.
Its technical details will be reported elsewhere~\cite{bulent}. 
In order to extend the SMF approach to the asymmetric system,
including the motion of the neck position is necessary.
%
%\begin{eqnarray}
% &&f(x,p_x ,t) \nonumber \\
%&=& \iint dydz\int ds_x\exp\left( - \frac{i}{\hbar}p_x s_x\right) \nonumber \\
%&&{}\times  \sum_{i\sigma\tau}
%\Phi_{i \sigma\tau}^\ast \left(x + \frac{s_x}{2},y,z,t\right)
%n_{i}^{\sigma\tau} \Phi_{i \sigma\tau}\left(x - \frac{s_x}{2},y,z,t\right).
%\nonumber \\
%\label{eq:wigner}
%\end{eqnarray}
%
Following the nucleon exchange picture~\cite{feldmeier87},
the drift term can be estimated through the semiclassical expression
of the SMF approach:
\begin{eqnarray}
v_{A}(t) &=& \int\frac{dp_x }{2\pi \hbar}
\frac{\vert p_x-p_0 \vert}{m} \nonumber \\
&&{}\times\sum_{\sigma\tau} \left[f_P^{\sigma\tau}(x_0,p_x ,t)
-f_T^{\sigma\tau}(x_0,p_x,t)\right],
\label{eq:drift}\end{eqnarray}
where $p_0/m$ is the velocity of the neck position.
On the other hand, the drift term in TDHF is given with the help 
of the Wigner function as
\begin{eqnarray}
v_A(t)= -\int \frac{dp_x }{2\pi \hbar}\frac{ p_x-p_0}{m}f(x_0,p_x ,t),
\label{eq:driftTDHF}
\end{eqnarray}
where the Wigner function is given from Eq.~(\ref{eq:wignerPT}) by
\begin{eqnarray}
f(x,p_x ,t)=\sum_{\sigma\tau}\left[f_P^{\sigma\tau}(x,p_x ,t)
+f_T^{\sigma\tau}(x,p_x ,t)\right].
\label{eq:wigner}\end{eqnarray}
The expression of Eq.~(\ref{eq:drift}) can be seen as a semiclassical 
version of Eq.~(\ref{eq:driftTDHF}). 
We have recently examined its validity 
by comparing the average number of nucleon transfer,
\begin{equation}
A_T(t=\infty)- A_T(t=0) = \int_0^\infty dt\, v_A(t),
\label{eq:integ-drift}
\end{equation}
where $A_T(t)$ denotes the mass number of the target-like fragment at time $t$.

\begin{figure}[bthp]
\begin{center}\leavevmode
%\resizebox{\columnwidth}{!}{%
  \includegraphics[width=\columnwidth,clip]{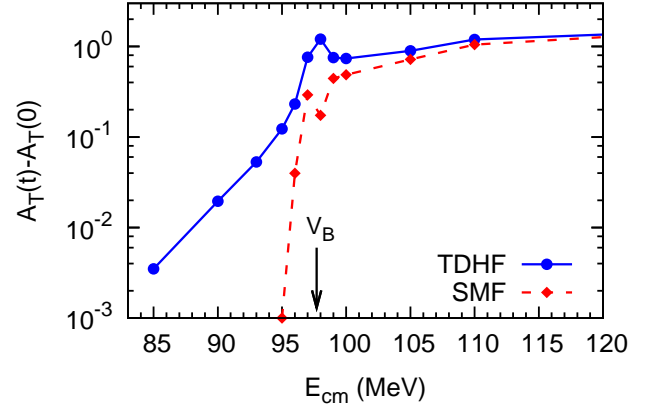} 
\caption{
The average number of nucleon transfer from $^{40}$Ca to $^{90}$Zr evaluated 
from TDHF (solid line) and from the semiclassical expression of SMF 
(dashed line) are shown as a function of center-of-mass energy for 
the head-on $^{40}$Ca+$^{90}$Zr reaction.
The Coulomb barrier $V_B$ is indicated as the arrow.
}
\label{fig:driftCaZr}
\end{center}
\end{figure}

Figure~\ref{fig:driftCaZr} shows the result of Eq.~(\ref{eq:integ-drift})
obtained by TDHF and by the semiclassical expression of SMF 
as a function of center-of-mass 
energy for the head-on $^{40}$Ca+$^{90}$Zr reaction. 
The semiclassical expression which has been used at energies 
above the Coulomb barrier leads to values consistent with its fully
quantal counterpart.
However, it clearly breaks down at energies below the Coulomb barrier.
This is due to the fact that the effect of nucleon tunneling 
is not treated well in the semiclassical expression below the Coulomb barrier,
underlying the necessity of quantal description of nucleon transfer.

\section{Conclusion}

Mean-field dynamics and mean-field fluctuations using microscopic time-dependent models
are discussed in the context of low energy nuclear reactions.
We have shown that the TDHF theory gives precise values of nucleus-nucleus
potential and a universal behavior of energy dissipation.
By projecting the SMF equation on the relative distance of colliding nuclei
and the mass number of the projectile-like nucleus,
we extract the corresponding diffusion coefficients.
%The expression of the diffusion coefficient has a similar structure with those familiar from the phenomenological nucleon exchange model. 
%Comparison between the calculated variance and the number of exchanged nucleon 
%supports a strong confirmation for the fact that the SMF approach provides 
%a realistic description of dissipation and fluctuation dynamics
%at low energies. The stochastic extension of the mean-field theory provides 
%a practical solution to the estimate of fluctuations of observables at low energies.  
%
%We have shown that fluctuations comparable with macroscopic models are recovered using SMF. 
We have shown that the SMF approach correctly describes 
the mass variance of final fragments in transfer reactions
at energies near the Coulomb barrier.
This gives a practical solution to properly describe mean-field fluctuations
of one-body observables at low energies.
%
%
%As a continuation of this work in the SMF approach,
%a work on an application to non-symmetric heavy-ion reactions
%near the Coulomb barrier is in progress~\cite{bulent}.

\begin{acknowledgement}
 This work is supported in part by US DOE Grant DE-FG05-89ER40530.
\end{acknowledgement}

\end{document}